# NSF Broadband Research 2020 Report

Henning Schulzrinne and Marie-José Montpetit

June 26, 2021

# Acknowledgements

The authors would like to thank all the participants of the three workshops for their enthusiasm, their commitment, and their important contributions. The breakout sessions scribes, in particular, deserve special thanks for their detailed notes that made this report possible. The workshop and report were supported by the National Science Foundation under Grant No. CNS 20-38333.

# Introduction

The internet has become a critical communications infrastructure, and access is among the "assets, systems, and networks, whether physical or virtual, [that] are considered so vital to the United States that their incapacitation or destruction would have a debilitating effect on security, national economic security, national public health or safety, or any combination thereof." [CISA] But the internet is more than an issue for the nation as a whole. Internet access affects the security, health, safety, and opportunities in life for individuals and communities, and the economic vitality of businesses everywhere.

On the one hand, the COVID-19 pandemic has revealed the success of broadband access in allowing society to function, even during lockdowns. On the other hand, the pandemic has exposed weak, unreliable, or even nonexistent, broadband access and usability in many areas and for many individuals, including especially rural residents and children in many school districts, urban, and rural, as well as a compelling need for universal reach and affordability. It was thus timely that in November 2020, the National Science Foundation (NSF) sponsored a series of workshops to identify new research areas to drive the broadband NSF agenda for the next five years. This request followed similar initiatives during the past twenty years, the most recent in 2016.

This *Broadband Research Workshop Report of 2021* discusses the research questions and challenges that need to be addressed to provide robust, affordable, and meaningful broadband access to every resident of the United States.

The workshops took place, as video conferences, between November 4 and November 18, 2020. They were loosely structured around the themes of technology, economics and digital inclusion. The meetings centered on a series of questions for the participants, briefly summarized as the *which*, *what* and *how* of broadband: *which* technologies, economic tools, and social support require more research to facilitate broadband access in the next five years, *what* is the research that will clarify the economic and societal factors influencing this deployment and adoption, and *how* to deploy new broadband networks and ensure access by everyone who needs it.

The online workshops focused on technology, economics, and digital inclusion. Each was spread over two days to accommodate the different time zones of the participants. Participants represented academia, nonprofit research organizations, and NGOs along with some industrial and operator researchers. They were faculty, senior researchers, and advocates or even activists, many with international reputations and distinguished research credentials. There was a wide geographical distribution in the US to include voices from across the country with some participation from Europe and Canada. Over 65 participants, listed in the Appendix, accepted the invitation. This report updates and adds material and references published since the conclusion of the workshops. References are not meant to be a survey of the large body of related research, but rather provide examples of relevant research or advocacy, with no attempt to present the most informative or foundational work.

The workshop and report writing period spanned a time of unprecedented interest at all levels of government and across many civic institutions in addressing the lack of broadband access, adoption and use. Considerable public investments are already being made in broadband infrastructure, enhanced emergency broadband subsidies, and digital inclusion efforts in state and local governments. Billions of dollars have been expended in the various COVID relief bills; any congressional infrastructure bill appears likely to include additional resources for broadband. To make these investments effective, we need data, research directions and research funding to evaluate what works and the social impact of these investments. Thus, addressing the research questions posed here not only provides a better understanding of the technology and its use, but also is likely to have a broader impact in informing public policy discussions.

The research discussed here also addresses, in part, one of the questions posed by Pres. Biden to Dr. Lander, the President's Science Advisor and Director of the Office of Science and Technology Policy: *How can breakthroughs in science and technology create powerful new solutions to address climate change—propelling market-driven change, jump-starting economic growth, improving health, and growing jobs, especially in communities that have been left behind?*[1]

---

[1] January 2021; available at
https://www.whitehouse.gov/briefing-room/statements-releases/2021/01/20/a-letter-to-dr-eric-s-lander-the-presidents-science-advisor-and-nominee-as-director-of-the-office-of-science-and-technology-policy/

# Motivation

Broadband internet access has become a core infrastructure, comparable in importance to water, electricity, and transportation. Few people would purchase a home that does not offer running water, reliable electricity and a road to get there – and increasingly, few would want to live permanently in a home without fast, reliable and affordable internet connectivity. This transition from a valuable, but somewhat secondary, infrastructure to a crucial utility, largely during the last decade, has made researching broadband connectivity ever more relevant. While the United States and other high-income countries have now deployed some kind of internet access to above 90% of households and nearly 100% of young adults access the internet by various means (typically by smartphone), the adoption and availability of key internet applications with societal benefits, such as remote work, telehealth and distance education, remains much lower, particularly for households with lower income and those living in rural areas." A [recent Harris Poll](https://4-h.org/media/new-survey-finds-teens-without-access-to-high-speed-internet-are-less-likely-to-believe-they-can-achieve-the-american-dream/)[2] commissioned by National 4-H Council and Microsoft found that 21% of teens do not have internet access at home but instead rely on schools, libraries, and other public places to get access.

The COVID-19 pandemic has exposed both the potential of broadband and the remaining challenges. Residential internet access has made it possible for many white-collar employees to work from home, has allowed potentially infected individuals to consult with healthcare professionals without endangering them, and has allowed some K-12 and college students to continue their education, even if diminished, during the spring of 2020. But it has also starkly illustrated the inequities in access, where some students could approximate their classroom experience by video conference and engaging with rich instructional materials, while others had to rely on school buses providing Wi-Fi and pick up work sheets while sitting in a library parking lot, if they could get broadband at all.

The primary focus of this workshop report is the "internet (broadband) as infrastructure," i.e., the concerns that arise trying to make broadband universal, affordable, and reliable, as well as to provide an infrastructure that enables rather than constrains novel applications and improvements of existing key applications. Architecturally, the emphasis is on network access, i.e., the so-called last mile, in all modalities, and key related infrastructure such as middle mile, internet exchanges (IXPs) and edge cloud computing.

However, this does not mean that discussions reported here are restricted to broadband access infrastructure in isolation. An important dimension of infrastructure is the connection of technical access with educational resources, housing, job requirements, small business needs, individual skills, and other important aspects of the policy and societal context. Of interest are the links between digital inclusion and improved social outcomes, including health, educational attainment, social mobility, and financial literacy.

---

[2] https://4-h.org/media/new-survey-finds-teens-without-access-to-high-speed-internet-are-less-likely-to-believe-they-can-achieve-the-american-dream/

Since the last [Report](#) was published in January 2017, concerns about internet applications, privacy, security and dominance by a small number of global enterprises have consumed a larger share of attention from researchers and the public. These topics, however, require different research approaches than the more infrastructure-focused research discussed here, are addressed by a different research community and will likely also have a different audience of "research consumers" in the policy analysis community, governments, and civic society. Thus, while we anticipate that concerns about privacy and misinformation, for example, will enter into discussions about non-adoption, they are beyond the scope of this effort.

The 2017 report, including the [2016 "Broadband 2021"](#) workshop report, still raises many research issues that deserve attention, and will likely continue to be relevant until we have universal availability and adoption, but they seem insufficiently precise and actionable, in our view. Indeed, in many cases, one could replace "broadband systems," for example, with "transportation system" or "water system" and ask similar questions. Thus, one goal of the workshop was to dig one layer deeper. As an example, Technology and Engineering Economics RQ1 (page 8) asked "What factors should be considered in engineering-economic models of broadband systems that employ novel technical approaches?" A more focused question might ask: "What factors currently drive capital and operational costs in rural and urban 5G and fiber networks, what technologies and approaches have shown promise to reduce those expenses and enhance capabilities and services, what has prevented their adoption so far and what experiments or data are needed to validate these approaches?"

Given the audience and to maintain focus, this report primarily considers broadband in the context of the United States, including its particular social, economic, topographical and regulatory conditions. However, since many of the problems, such as reaching rural areas, as well as research questions, data issues, and approaches are similar, other viewpoints are incorporated and there are references throughout the report to projects and programs insights from the European Union, the United Kingdom, Australia, Brazil, and many African countries that have comparable broadband concerns and strong research efforts. The participants were made aware the the target audience would be:

- Federal funding agencies that will require information on why they should create new programs or solicit specific kinds of research;

- Mission-specific agencies (at the federal and state levels) who will need to know why they should support this research in their program design and operation;

- The research community-at-large that may be interested in topics that deserve more attention and the set of tools and data sets that are under-utilized and could be helpful in their work.

## Foundational Questions

A list of questions were used to initiate the discussion in the workshop, even though we were not able to address, let alone answer, all of them:

- Retrospective:

- What have we learned in the last five years?
- What has changed since 2016?
- What assumptions and projections were less accurate and why?
- Which (major) policy ideas over-promised and under-delivered?
- What major research challenges remain open and why?
- To what extent has COVID-19 changed our understanding of broadband usage, deployment and impact in the short and long terms?
- Are there examples of research (and papers) that illustrate the answer to these questions?
- Prospective:
    - What should be done next?
    - Key research topics and questions, prioritized:
        - Why does this matter to the research community?
        - What does it help us do or understand better?
        - Not only a random list of worthy ideas, but a focus on major challenges
        - State goals where appropriate ("for example: increase the efficiency of federal subsidies by 30%")
    - Data needed and why:
        - What hurdles complicate getting that data?
        - How can agencies coordinate to improve data access?
        - How can they design programs to measure while doing?
    - Other resources
        - e.g. open source software and hardware, testbeds or pilots?

These questions were also used to request position papers to propose ideas, define concepts and point to recent related work before the workshops. The workshop organizers received about 40 position papers. Twenty-eight of these papers are publicly available at https://edas.info/p27906. Other background information about the workshop can be found at

https://sites.google.com/view/broadbandresearch2020/home

Each workshop day included breakout panels to address topics specific to the weekly theme and plenary discussions on the second day allowed the scribes of each session to present the salient discussions of each subgroup and attempt to answer the questions above.

The results of all these discussions are presented in this report. The format of the report roughly follows the workshop timeline but some cross-cutting topics were also identified and are presented first.

# Goals

The overall goal of this report is to describe the key research questions for broadband for the next roughly five years, why they are important to reaching societal goals and what the federal government, non-governmental organizations and academic institutions can do to advance these research objectives.

This report summarizes the discussions on the progress in broadband research, deployment and use since the last report. It highlights continuing and emerging research issues in broadband technology and deployment, the economics of broadband and digital inclusion, recognizing that these issues are intertwined.

The report, building on the [2016 "Broadband 2021"](#) approach, aims to

- identify important and timely research areas, in engineering, economics and social sciences;

- describe existing and missing collections of data that enable and ground that research;

- explore how research efforts can be better coordinated across the federal government, the academic community, and the broader broadband research community.

The report also provides recommendations, such as new or modified research funding, research accompanying federal broadband deployment and adoption efforts, and the necessary infrastructure (e.g., data or testbeds) to improve research quality and impact.

The workshop and report were guided by a set of guiding operating principles:

- **Actionable recommendations:** The workshop participants were asked to formulate actionable and verifiable goals, with sufficient concreteness that their completion can be determined factually. For example, rather than asking for generic "better data," the report might identify specific data sources and who would need to gather the data or make it available.

- **Prioritize research goals:** It is tempting to simply enumerate goals and research issues, but in an environment with severely constrained resources, the workshop participants were asked to rank or weight objectives by their impact and resource needs, distinguishing between "nice to know" and "critical for good policy making," for example.

- **Identify necessary resources:** The workshop participants were also asked to clearly describe the nature of the resources, e.g., funding, staff, legal or regulatory authority, necessary to implement goals and objectives. As the 2017 report noted, in some cases, having institutional, long-term resources and data collections is particularly important, and may take precedence over granularity.

- **Reflection on audience:** Broadband research has the potential to influence policy, such as the eight-billion dollar annual expenditures for the FCC universal service program. Other research might guide practitioners in digital inclusion efforts or help prioritize research funding. Thus, the research issues should clearly identify the audience – who needs this information, how timely and frequent does it need to be updated, and what does the likely "consumer" of the research need to know to help them?

- **Inclusive:** As infrastructure, broadband touches everybody, but particular groups have been particularly affected by lack of access or use, including people with disabilities, Native Americans, low-income households, people of color and seniors. The workshop will strive to incorporate voices from communities representing these groups, to help identify research needs and challenges.

# Broadband Research

## What Differentiates Broadband Research?

There has been a thriving and substantial network *engineering* research community in the United States since the 1980s, largely financially supported by the NSF CISE directorate and, to a lesser extent in recent years, DARPA and the telecommunications and information technology industry. Similarly, research focused on network technologies and applications has been a dominant theme of the various European research programs, such as the Framework Programmes for Research and Technological Development (1984 through the present).

Here, however, we focus on broadband research as research *that studies and furthers the universal availability, meaningful adoption and positive societal impact of broadband internet access*. While engineering research justifiably focuses on improving the technical performance of networks and explores new applications, broadband research draws on both the engineering disciplines and the social sciences, in close interaction.

## Cross-Cutting Themes

Throughout the workshop, a small number of themes made repeated appearances, regardless of the specific research question or area, in particular the importance of data sets and metrics.

The pandemic revealed persistent gaps in broadband access and use, discussed in depth during the workshops. These gaps create "broadband deserts" that often coincide with healthcare deserts and fresh food deserts. As discussed later in the report, these broadband deserts affect people's access to health care and job opportunities, government services (including vaccine access), and their children's learning and development, with long-term consequences. Gaps in access and use have been identified through a variety of means, including investigative journalism, surveys by nonprofit research groups, research based on government data sources and interview studies by academic researchers. The result has been a growing consensus that broadband access can mean the difference between thriving and languishing. To build on these observations toward generalizable knowledge will require systematic, representative surveys, careful analytics, and computational social science, and systematic qualitative research to better understand how broadband has the potential to improve all lives and why inequities persist.

## Broadband Data Sets

Data for enabling broadband research, its acquisition and interpretation was a dominant theme across all three workshops. Here, data sets primarily describe the deployment, performance, and impact of broadband networks, as well as datasets that provide demographic, topological and economic context. The emphasis on data reflecting observations of broadband deployment, usage and impact is probably one of the distinguishing features of broadband research compared to, say, engineering-focused research.

While many research projects gather data for their own use, and sometimes make that data available to other researchers, the workshop emphasized the need for common, shared and maintained data sets that can facilitate research across many different projects, without burdening each project with the substantial cost of acquiring, storing, cleaning and processing the data. Longitudinal data, capturing the historical evolution of a metric, is often particularly useful to test models and evaluate the impact of broadband-related public policies. The lack of such data sets also makes it nearly impossible to reproduce and extend earlier research results, e.g., by extending the analysis further into the past or beyond the original time horizon of the research.

All three workshops recognized the role of NSF as well as the US Census Bureau, the NTIA and the Federal Communications Commission (FCC), along with the Universal Service Administrative Company (USAC), and possibly commercial entities such as network measurement companies, telecommunication carriers and their trade associations, play in creating and maintaining the data necessary to advance broadband research, albeit with different roles and responsibilities. In addition, agencies that fund broadband deployments, such as the Department of Agriculture, through its ReConnect program at the federal level, may already or could generate valuable data on broadband deployment and use. The National Science Foundation has recently created a broadband deployment pilot program, Project OVERCOME[3], that combines research along with deployment and adoption activities.

Broadly speaking, there are four kinds of broadband-related data sets:

- *Industry data collections* such as the current FCC Form 477 data, where providers are legally required to provide information, but not all such data is made publicly available;
- *Administrative data* such as information about locations subsidized by the FCC universal service high-cost programs;

---
[3] See https://www.nsf.gov/awardsearch/showAward?AWD_ID=2044448

- *Survey data* on broadband usage and availability, such as data gathered by the Census Bureau, NTIA, Pew Research and other organizations;
- *Measurement data* that directly observes the performance or usage of access networks.

We discuss each of these in more detail in the related section, depending on their use.

Even where data is made available, it is usually in a format that requires extensive data import effort, such as very large CSV files. Precise definitions of the data columns are often either missing completely or buried in agency reports, making it far too easy for researchers to inadvertently misinterpret data. In some cases, definitions differ across agencies, making it difficult to integrate data sets.

Some agency open data efforts provide web-based query interfaces, but they are typically unwieldy, not suitable for large datasets and do not allow cross-table joins, e.g., to combine broadband and demographic data. Others provide map-based interfaces to broadband deployment data, such as the national broadband map, but these are mostly useful for quick visualizations for the public or policy staff and generally not particularly helpful to researchers (but do help with creating presentations for students and panels!).

Longitudinal and panel datasets, i.e., datasets that cover the same variables for the same geographic region or set of individuals, are particularly valuable for researchers, as they allow for difference-in-differences studies. Unlike most university research groups, federal agencies have the institutional capacity and mission to create, maintain and document such datasets.

**Recommendation:** All agencies producing broadband-related data should regularly seek the input of broadband researchers to maximize the utility of existing or planned datasets and to identify gaps in data or definitions.

**Recommendation:** All agencies with broadband-relevant datasets should allow researchers to easily apply for access to restricted data, with appropriate protections of confidentiality, and consider use of data enclaves.

**Recommendation:** All agencies should coordinate broadband-related definitions and data gathering activities, e.g., through the Broadband Research Council. The Council or one of the federal agencies with mission focus on broadband should publish and maintain a survey of key broadband data resources suitable for researchers, including known limitations or caveats.

**Recommendation:** Broadband-related projects, whether supporting infrastructure or adoption, should strive to collect data concurrently with these projects and make provisions to collect impact-data beyond the duration of the broadband projects.

**Recommendation:** NSF or NTIA should provide or fund a data hosting infrastructure that makes broadband-related data sets available for queries at scale, e.g., through an SQL-like interface.

## Metrics

Metrics are systematic measures used to assess a goal or extract information from data. The need for new and better metrics was a recurring theme across all three workshops. We need metrics to better inform policy, to demonstrate the most effective and efficient approaches to expanding broadband access and use. For instance, as wireless technologies, both fixed and mobile, are becoming a key means of accessing the internet, we need metrics that reflect their more variable performance in time and space. We also need metrics that identify broadband adoption as more than a binary variable, reflecting different intensity and extensivity of use. Impact metrics are needed to assess digital literacy, job-specific broadband requirements, effects on job performance, consumer behavior, access to health care, and education, among others. Having consistent and well-defined metrics makes it possible to aggregate smaller-scale studies into more robustly-supported results and insights.

**RQ:** What are key metrics for broadband quality and quality-of-experience for both fixed and mobile internet access?

**RQ:** What are well-defined metrics that describe the extensivity and intensivity of broadband adoption and use?

**RQ:** Can we measure digital literacy and quantify the digital skills needed for broadband use of particular interest?

## Human-Centered Research

Participants in all the workshops emphasized the need for human-centered broadband research that includes an assessment of people's social, physical, economic, and cultural context. Human-centered research would include, for example, improved assessments of the broadband needs of different communities (for example, in the provision of local government services), the resilience of different communities and groups during crises and disasters, the impact of alternative ways of making people aware of broadband subsidies, or even the usability of advanced network management for small operators.

With much of what we know about the "user" side of broadband dependent on spotty data and anecdotal reports, it is difficult to know how much of the apparent digital divide is driven by the lack of broadband access, by affordability, or by people's disinterest or disinclination to use broadband even if the economic costs are minimal or absent. Indeed, despite numerous attempts to answer this question for the last twenty years, the debate does not seem to abate.[4]

Effective use of broadband may also be hindered by poor usability, whether for people with motion or sensory impairments or the often frustrating experience when internet access or internet services fail in opaque ways. Particularly for traditional "desktop" home computing, security configuration, credential (password) management and fault recovery can challenge all but the most technically-skilled users.

A usability gap remains between traditional home computing and mobile devices. Most, for example, manage to set up and use key applications on a smartphone, but struggle to set up or repair a home Wi-Fi network or to effectively use two-factor authentication.

---

[4] This may also be partially driven by the concern that certain conclusions may lead to questions about broadband industry structures and competition.

Smartphones and tablets have moved beyond being only passive one-way devices for consumption; they are often used to create audio and video content, for example, even as they remain less suited for more complex content creation tasks. Due to lower cost, mobility and higher usability, these devices have become the dominant means of internet access for many. In the future, smart speakers and embedded devices may offer an additional modality for user access.

But government and corporate websites were often designed for large screens and keyboard interaction, or provide limited accessibility to users relying on, say, screen readers or for whom English is not the first language, making it difficult for all users to make full use of these resources. Research on broadband usability, such as that supported by NSF's Human Centered Computing program or through research in usable security, could result in the design of improved universal design for network technology and software interfaces, through guidelines, tests, education and training. (For example, many computer science students never learn about universal design.[5]) However, we lack a good grasp of which elements of universal design are particularly helpful to foster effective adoption of broadband and how to measure success, e.g., as part of product reviews or certifications. The need for improved security poses increasing usability challenges for many, possibly raising barriers to use.

**RQ:** How does the lack of usability and failures to observe best practices of universal design affect broadband adoption and effective use?

**RQ:** How can principles of universal design be applied to the setup, configuration, fault diagnosis and security-related interactions of home broadband?

**RQ:** How can we measure and improve the usability of home broadband systems and key applications? How can we translate research insights into deployed systems?

## Broadband Technology

The technologies used to provide internet access influence the cost, performance, reliability and security of broadband. Both the overall system architecture and component technologies matter; given their large scale and long component lifetimes, change in both often takes significant time to move from research to products to deployment. While the options for internet access have not changed fundamentally in the last five years, the number of homes passed by fiber has roughly doubled,[6] while fixed wireless and low-earth orbiting (LEO) satellite constellations have attracted increased attention. Below, we discuss research issues for both the overall network architecture and key access modalities.

---

[5] "Universal Design is the design and composition of an environment so that it can be accessed, understood and used to the greatest extent possible by all people regardless of their age, size, ability or disability." (http://universaldesign.ie/What-is-Universal-Design/)

[6] https://www.bbcmag.com/multifamily-broadband/fiber-trends-what-2021-promises-for-the-broadband-industry

While there is much discussion about what speeds networks should have, and whether the current level of speed asymmetry in networks other than fiber-to-the-premises networks is sufficient or hinders the development of new applications, a key consideration will be supporting increased network load, i.e., per-household gigabytes per month. For example, the median household served by Comcast increased its data usage from 88 GB/month in December 2016 to 346 GB/month in December 2020.[7] The average (mean) usage is significantly higher, predicted to be around 660-650 GB/month[8] at the end of 2021, with per-year growth of 25% expected to continue. Thus, a key metric of success, particularly for fixed wireless, cellular and satellite networks, will be increasing overall capacity and reducing the dollar-per-Gigabyte ratio, not just or primarily to increase peak speeds.

## Network Architecture

Software-defined networks[9] have become a key network research topic, but the efforts largely assume that the programming is left to the owner of the infrastructure. SDNs have been most successful within datacenter networks and may make inroads in some carrier networks as a way to replace expensive routers with "white box" components. But just as computing and storage hardware has been made available as a leasable service via cloud computing providers, it may be possible to make key network components available to users, both internet service providers and end users. Unlike today's provisioning-oriented model that is provider specific and often semi-manual, allocating various elements, from "layer 0" resources such as ducts, pole space and tower attachments, to spectrum and radio access networks. Nascent efforts exist for dynamic spectrum access, but a more integrated approach may make provisioning access networks more efficient and possibly increase competition.

Starting in 2010, NSF funded a large-scale effort ("Future Internet Architecture") exploring four different directions for extending or replacing the existing internet architecture. Much has changed since those projects have been conceived, both in terms of dominant network use cases, technical and policy challenges, as well as alternative, industry-driven approaches that leaned more heavily on application-layer and transport-layer approaches such as CDNs and HTTP/2 and HTTP/3. Also, our understanding of the costs and operational challenges of networks has improved significantly since 2010, as well as the difficulties of transitioning to new architectures, thus inviting a new look at how we can significantly reduce operational costs, improve deployability and increase resilience of access networks. For example, compared to a decade ago, production networks today span a much larger range of technologies, from low-power wide area networks (NB-IoT and LoRaWAN) to Wi-Fi, fixed wireless, fiber, hybrid fiber-coax (HFC), legacy DSL, and low-earth and geosynchronous satellites.

---

[7] https://www.xfinity.com/support/articles/data-usage-average-network-usage

[8] https://www.nexttv.com/news/average-monthly-us-broadband-usage-could-reach-650-gb-in-2021

[9] "In the SDN architecture, the control and data planes are decoupled, network intelligence and state are logically centralized, and the underlying network infrastructure is abstracted from the applications." (Open Networking Foundation at
https://opennetworking.org/sdn-resources/whitepapers/software-defined-networking-the-new-norm-for-networks/)

A small but active research community is investigating alternative architectures that deviate significantly from the datagram internet protocol model, e.g., using named data networks, but they have so far not made significant inroads into production networks.

Since the 2016 report, network automation has become an industry focus, as carriers recognize that the long-run costs of networks are dominated by operations, i.e., configuring networks, adding and upgrading customers and diagnosing faults.

**RQ:** How can network infrastructure be offered as a service ("NaaS")? Which aspects of the infrastructure components, including spectrum and radio access networks, should be provisionable, programmable, by whom and how?

**RQ:** What aspects of network operations can be automated, including for smaller operators? Can some operations be effectively delegated to specialized services, such as security or QoE monitoring?

**RQ:** Can network slicing be extended to networks other than 5G cellular networks, can it span multiple independently-operated networks and can slicing enable new applications or use cases? Can we characterize the scale and nature of the use cases? What can we learn from earlier, less-than-successful, attempts to deploy quality-of-service technologies?

**RQ:** What network applications can benefit from higher upload speeds, whether on average or in relatively short bursts, e.g., for content upload?

**RQ:** Can networks be designed so that they support varying ratios of download-to-upload speed or capacity, either to reflect long-term changes in usage or to accommodate individual household or business needs?

**RQ:** What are the architectural challenges of access networks that combine wired, fixed wireless, cellular and satellite technologies?

**RQ:** Can alternative network architectures including the use of data centric and compute-centric architectures significantly and measurably increase reliability, cost effectiveness and security for current and predicted network use cases?

## Wired Access

Almost all fixed networks being built today are "fiber plus", whether they terminate fiber at the home (known as fiber-to-the-home (FTTH) or fiber-to-the-premises (FTTP)) and connect devices via Wi-Fi or, as below, fiber to the node for some DSL and all hybrid-fiber-coax ("cable") networks or fiber to the tower for fixed and mobile wireless. These networks are being built by both traditional carriers, large and small, and new entrants, from municipalities like Ft. Collins, Colorado, to small and large electric utilities and cooperatives. While fiber technology is considered mature and largely the remit of industrial research and standardization, larger architectural and protocol questions remain.

**RQ:** How can the design of passive optical network (PON) fiber networks be optimized for cost, resiliency and future expansion? Can open-source design tools facilitate planning and evaluation of such networks by potential new entrants and agencies that fund broadband deployment and evaluate grant applications?

**RQ:** Can robots play a role in accelerating the deployment of fiber networks, both aerial and underground?

**RQ:** What are the principles and design guidelines for optimizing hybrid fiber/fixed-wireless networks, both in urban and lower-density areas?

**RQ:** What are the primary technical challenges to operate such networks, particularly for smaller network operators?

Unfortunately, it is difficult for academic researchers to get insights into the operational and architectural challenges of building such networks. While there are a number of NSF-sponsored wireless testbeds as part of the PAWR initiative, building a lab-scale version of an FTTH network remains well beyond the reach of most research groups. While standard network textbooks explain DSL and HFC networks, they do not typically cover passive optical network architectures.

## Wireless Access

In the last five years, fixed wireless networks have assumed an increasingly important role in providing access, not just in rural areas as WISPs but also as an overlay and competitive option in urban areas. These are no longer the city Wi-Fi networks that were popular earlier,[10] but rather often-proprietary fixed wireless networks relying on consumers installing outdoor antennas pointing at a base station. However, estimates of how many households obtain internet access via such networks differ widely, from 1.5 million[11] to 4 million[12] in 2018.

The heated discussion after the Rural Digital Opportunity Fund (RDOF) auctions have illustrated that the performance of fixed wireless networks at scale is still a contested question. Few real-world measurements by parties without a stake in the outcome cover a diverse range of topologies and technologies, making it difficult for agencies funding broadband build-outs to predict whether and which technologies are likely to meet expectations. The existing NSF PAWR networks focus on programmable and novel physical layer technologies and thus have a longer time horizon, rather than providing access to commercial (non-Wi-Fi) technologies likely to be deployed in the next five years.

While the Internet of Things (IoT) was mentioned in the 2016 report, the emphasis was on their impact on middle-mile and access networks. However, IoT has now separated into two domains: in-building fixed-location devices, largely connected by Wi-Fi and served by existing access networks without much impact on their traffic load, and outdoor or mobile applications. The number of indoor and in-home devices, including devices largely unknown five years ago such as smart speakers, has increased dramatically, while the number of connected outdoor devices has fallen short of expectations. It had generally been assumed that outdoor IoT applications, such as agricultural sensors or vehicle tracking, would rely on cellular network infrastructure, such as LTE-M or NB-IoT. But specialized low-bandwidth networks operating in unlicensed bands, such as LoRaWAN or DECT-2020 NR, may be able to cover large geographical areas without permanent residential or commercial structures, i.e., where deploying cellular or fixed-wireless networks is economically unattractive.

---

[10] See, for example, "Municipal WiFi: The Coda", *Journal of Urban Technology,* 2010.
[11] FCC, "Internet Access Services: Status as of December 31, 2018", September 2020 (https://docs.fcc.gov/public/attachments/DOC-366980A1.pdf)
[12] Carmel Group, "The 2021 Fixed Wireless and Hybrid ISP Industry Report," (https://www.wispa.org/docs/2021_WISPA_Report_FINAL.pdf). Note that the subscriber figures in the report include hybrid ISPs, i.e., those that provide service both by fiber and other wired technologies, as well as wireless.

**RQ:** How can we further extend the performance frontier of maximizing distance and bandwidth, while minimizing latency? In particular, self-configuring highly-directional links that adapt to their propagation environment may allow extending the reach of networks across sparsely populated rural areas.

**RQ:** How can we better estimate the actual performance of fixed wireless networks under real-world conditions?

**RQ:** Can 5G and future cellular radio technologies be repurposed as physical and link layer components for fixed wireless networks?

**RQ:** What can 6G improve over 5G in terms of data driven approaches and better use of networking resources beyond high bandwidth and low delay?

**RQ:** Can low-bandwidth wireless networks offer internet of things or emergency notification connectivity in areas with very low population density or no permanent settlements? For each architecture or technology, what would it take to provide connectivity across every square mile or every road mile of the United States?

**RQ:** How can IoT networks be better organized to profit from broadband connectivity in sharing and orchestrating resources, while protecting user privacy?

**Recommendation:** NSF or other agencies should set up a real-world testbed, possibly as part of an existing testbed, that provides access to commercially-available fixed wireless equipment across urban, suburban and rural terrain, including forested and mountainous locations. They should fund work on neutral, reproducible and credible performance evaluations of such technologies.

**Recommendation:** One or more fiber or hybrid fiber/wireless networks, e.g., operated by a municipality or cooperative, should allow access to researchers to better understand planning and deployment challenges, network operations and long-term sustainability.

## Other Access Modalities

Since the previous broadband research report, the dominant high-speed access technologies of FTTH, HFC and fixed wireless have been augmented with new approaches, primarily planned low-earth orbiting satellite constellations such as Starlink by SpaceX, Kuiper by Amazon and experiments with high-altitude platforms such as the (now abandoned) Google Loon project or UAVs. They may offer a different mix of cost, coverage and speed-of-deployment trade-offs, even if they may not reach the number of users served by more traditional access modalities. For example, they may allow rapid restoration of network access after natural disasters. Testbeds such as PAWR AERPAW may make experiments using UAVs as cellular platforms accessible to a wider range of researchers.

**RQ:** When, where and how are such new broadband access modalities likely to offer attractive alternatives or complements to more traditional access network technologies?

**RQ:** What are the capacities, lifetimes and observed reliability of these networks?

**RQ:** Intersatellite links may reduce costs and improve the performance of LEO constellations. How can their performance and reliability be improved?

**RQ:** Are there ways to do more with less bandwidth? Can applications adapt to function more effectively over connections with limited bandwidth or intermittent availability, preferably without manual tuning?

**Recommendation:** Testbeds should provide a wider range of airborne experimental platforms, such as tethered balloons or cubesats, as these will help explore additional temporary or long-term network connectivity options in challenged environments.

## Network Measurements

Network measurements play an important role for broadband research. Such measurements can help answer **research questions** such as:

**RQ:** Does the measured performance experienced by end users correspond to the advertised network bandwidth? If not, can performance impairments be traced to, say, home Wi-Fi, provider throttling, the last-mile network or provider interconnections?

**RQ:** How reliable are access networks and does reliability differ between technologies or types of providers? Does it appear to get better or worse? Are reliability problems typically brief or long-lasting, and are they complete outages or significant performance degradations or partial failures?

**RQ:** Does the usage profile of end users change as download and upload speeds increase? For example, are certain applications used more frequently in networks offering higher speeds or lower latency?

**RQ:** Does the subjective satisfaction of users change as network speeds and latencies improve? Does it matter if available speeds vary, e.g., at different times of the day?

Currently, there are three major, long-running network measurement efforts that go beyond individual academic research projects and that make data available to researchers or the public: the FCC Measuring Broadband America (MBA) project,[13] Ookla (speedtest.net) and the RIPE Atlas[14] project. In addition, some efforts have measured the actual download performance of cloud services; these measurements can provide some indication of the actually-subscribed speeds.

The FCC manages the Measuring Broadband America project, with several thousand volunteers across a varying set of major internet service providers hosting measurement boxes in their homes across the United States. (However, as of 2021, at least one major landline network provider does not currently participate and none of the consumer satellite operators do.) Since the measurement device is directly attached to the broadband or service gateway (home router), which may incorporate the cable modem, DSL modem or optical network terminal (ONT) functionality, the measurements are not distorted by impairments caused by home Wi-Fi. This effort was started in 2011 and thus provides good longitudinal data. Unlike all other public datasets, the subscriber service tier is validated and the rough geographic location of the subscriber is made available, thus providing some ability to analyze regional variations in performance while protecting the privacy of the volunteers. The FCC makes raw data available within a few months of collection, but reports and validated data appear at less predictable intervals. Among the three projects, MBA collects the most diverse set of metrics. The sample sizes do not correspond to the popularity of tiers or the subscriber counts of providers. Instead, the sample size is chosen so that each of the popular service tiers (e.g., 100 download, 10 Mb/s upload) has a sufficient number of samples for statistical validity. The data has limitations: It is insufficient to reliably identify any geographic variations in network performance and does not allow analysts to directly extrapolate to the overall performance of a provider or the broadband users generally, as the report and data do not reveal the share or number of users subscribing to a tier.

---

[13] https://www.fcc.gov/general/measuring-broadband-america
[14] https://atlas.ripe.net/

Network speed measurement results can vary significantly - the same household, at nearly the same time, could see performance differences of a factor of two across commonly-used consumer-focused test tools. In part, this may well be due to measurement methodology, e.g., the use of multiple parallel TCP connections or different in-net measurement endpoints. Even the FCC MBA measurements rely on proprietary measurement methods, rather than the IETF IPPM-standardized approaches, for example.

The current network measurement research usually targets large ISPs and thus ignores the needs of smaller service providers. Since rural homes are more likely to be served by such smaller providers, this leaves policy makers and researchers with limited information on how well these rural networks perform.

Only the FCC MBA measurements provide some indication of network reliability, an increasingly important consideration for many applications.

Most measurements are active, i.e., generate test traffic, and thus pose no particular privacy concerns. But it would also be helpful to gain insights into changing user behaviors and usage patterns, as happened during the pandemic. The FCC MBA infrastructure collects total traffic volume, but does not distinguish between traffic classes.

The model of deploying devices (RIPE Atlas, FCC MBA) or convenience samples does not scale well and makes it difficult to test anything except roundtrip latency and some approximation of download and upload speed. It would be more useful for researchers if each cable modem or optical network termination (ONT) device was able to report such measurements, and to make such measurements more programmable. The IETF LMAP[15] and the Broadband Forum TR-143 standard[16] are steps in that direction, albeit not widely deployed and more likely to be used for internal monitoring than research.

---

[15] https://datatracker.ietf.org/wg/lmap/about/
[16] https://www.broadband-forum.org/download/TR-143.pdf

While some of the aspects of measurement also require new metrics, it is also important to be able to link measurements at different layers and include application performance in order to fully characterise network performance.

**RQ:** Can we explain the differences in measured performance and are there opportunities for defining a robust performance measurement standard that makes such measurements reproducible? What aspects of measurements should be disclosed?

**RQ:** How can we scale up the measurement infrastructure, e.g., by integrating it into access routers or home devices, so that most households could be sampled continuously, for performance and availability?

**RQ:** How can we better estimate network reliability and temporal variations in performance?

**RQ:** How can we track the demands and usage of different classes of applications over time?

**Recommendation:** The FCC, NTIA or NSF should work with one or more ISPs or home router vendors to instrument a larger fraction of their network, allowing more granular measurements, and make it available to researchers, with appropriate safeguards.

## Network Management

Currently, it is difficult for researchers (and students) to get a close look at modern fiber-to-the-home networks or experiment with management systems and algorithms. In addition a large amount of network management research and efforts is devoted to large service providers. Thus, researchers often focus on aspects, such as BGP routing, that may or may not be the most difficult or resource-consuming management tasks. Also, many of the tools used by large, well-resourced carriers with domain experts in everything from security to managing interdomain routing may not be available to smaller organizations.

Recently, a lot of research has explored the use of machine learning (ML) to develop better network management algorithms that can be deployed across different types of operations. The move to cloud-based operations and SDN also simplifies that approach, as it makes more data available. Yet, the applicability of these tools to real networks remains to be proven.

**RQ:** What are common and challenging network management issues in operational networks, including those of smaller operators or those incorporating multiple different physical layers?

**RQ:** Can ML offer improved fault diagnosis and traffic engineering in real networks, including those of smaller operators?

**RQ:** Can we design simpler network management architectures and collections of protocols that can be implemented across a wider range of operators?

**RQ:** Can we create a "digital twin" of a realistic operator network and use that to improve and test network management protocols and algorithms?

## Other Observations in the Technology Group

Some attendees observed a disconnect between network research and solutions that support societal needs. For example, network research may focus on performance optimization, but may be less attuned to policy, economic and deployability constraints.

In general, there seems to be little systems, empirical or theoretical peer-reviewed research addressing *access* network technology issues. For example, the largest selective networking conference, *IEEE Infocom 2021,* contains no papers on access networks other than 5G.

As network technology matures, networking research questions for access and wide-area networks may increasingly resemble civil engineering, i.e., with an emphasis on evaluating and improving deployed systems, cost effectiveness, operational efficiency, reliability and resilience. However, most of the transportation, water and energy infrastructure studied by civil engineering researchers is public or at least highly-regulated, such as roads, railroads, air transportation and electric and water distribution. Almost all networks are run by private entities, which have historically been reluctant to share cost, operational and deployment data with researchers.

During the early days of the commercial internet, many carriers had significant research labs, often with good connections to the academic community, and with explicit efforts by NSF to foster these connections. Many carriers have downsized or abandoned internal outward-looking research and thus, academic research struggles to identify the operational needs of modern networks.

**<u>Recommendation:</u>** NSF should facilitate the information exchange between network practitioners and systems-oriented researchers, including smaller network operators.

# Economics

## Economics of Building, Maintaining and Operating Networks

From the 1980s through 2010 or so, the progress of broadband networks could be easily measured in the top speed they were able to deliver. Now, while speed and latency are still topics of basic research for cellular networks, economic considerations are likely to dominate as a concern for deploying, upgrading and operating networks. In other words, it is no longer a technical challenge to offer 1 Gb/s or 10 Gb/s to households, but rather an economic one.

As the technical capabilities of broadband technology have matured, insights into the cost of deploying and operating broadband networks have become more important for policy makers. Compared to physical transportation systems such as roads and rail or electric distribution systems, we have a poor understanding of the costs associated with building, maintaining and operating bit transport systems, i.e., broadband networks. Standard introductory, systems-oriented network textbooks do not discuss costs in any depth. But good data on both short-run and long-term costs of networks are crucial to make informed decisions on what type of rural network technologies should be funded. For example, we need to know if the initially lower cost of fixed wireless networks compared to fiber is outweighed by higher operational costs or earlier obsolescence. Similarly, the impact of 5G networks on rural areas largely depends on the cost of deploying small cells. While most investor-owned carriers only provide top-line investment and cost numbers, some smaller non-profit or governmental organizations may be more willing to share cost data, but gathering and contextualizing the cost data may well require significant effort.

Cost models for broadband seldom include lifecycle costs, including difficult-to-model user-borne costs for outages, service disruptions, and uneven coverage. Most models also do not quantify the impact of spectrum resources and licensing or auction models on the overall deployment costs.

**RQ:** What are the key cost components for building, maintaining and operating broadband networks in both urban and rural settings, for different technologies?

**RQ:** What are cost tradeoffs between increased use of edge CDNs and middle-mile deployments?

**RQ:** Does the nature of the service provider influence costs? For example, do electric co-ops or municipal electric utilities enjoy cost advantages because they can share poles and ducts or have lower customer acquisition or capital costs?

## Quality, Price, and Competition

The price of broadband services is often cited among the reasons that households do not subscribe. Broadband is often the largest utility expense for households, particularly renters. Policy makers are subject to very[17] different[18] interpretations of the price consumers pay relative to other countries, but everyone agrees that the data sets available to researchers make comparisons and analysis difficult. This is particularly difficult since U.S. broadband prices tend to vary significantly by region, with various bundles and service-related fees, and are often disclosed only during a phone call with a sales agent; some unknown number of households pay various time-limited promotional rates or benefit from income-restricted rates. For many years, there have been proposals for the FCC to collect rate information as part of its Form 477 process; currently, the FCC collects and publishes a sample of about 3,200 rates every year through its urban rate survey.[19] The Open Technology Institute collects plan information from providers in major metropolitan areas.[20] The Bureau of Labor Statistics gathers pricing information,[21] but does not provide a detailed breakdown of the nature of the plans or disclose the data points, making it difficult to judge how, for example, to adjust for quality, whether that is speed, latency or reliability, or whether the data captures the mix of pricing plans. Researchers often have to rely on web scraping[22],[23] to obtain pricing and offering data.

---

[17] Thomas Philippon, "How Expensive are U.S. Broadband and Wireless Services?" (April 29, 2021). Available at SSRN: https://ssrn.com/abstract=3836281 or http://dx.doi.org/10.2139/ssrn.3836281

[18] Mark Israel, Michael Katz, and Bryan Keating, "International broadband price comparisons tell us little about competition and do not justify broadband regulation," (May 2021), available at https://www.ncta.com/sites/default/files/2021-05/international-price-comparisons-paper-11-may-2021.pdf

[19] https://www.fcc.gov/economics-analytics/industry-analysis-division/urban-rate-survey-data-resources

[20] https://www.newamerica.org/oti/reports/cost-connectivity-2020/

[21] https://www.bls.gov/cpi/factsheets/telecommunications.htm

**RQ:** What are appropriate mechanisms for gathering broadband pricing information that allow in-depth quality and price analysis, both across time and different types of providers? Would a household panel, where members of the panel share broadband bills, be sustainable and credible?

**RQ:** How does retail competition at the household level affect consumer quality choices for broadband, such as speeds, bandwidth caps and bundles[24], retail prices, broadband adoption and network reliability?

## Measuring and Predicting Broadband Adoption and Use

Measuring broadband availability is hard, as we note below; measuring broadband adoption is even more difficult. Both the FCC and the Census collect broadband usage (subscriber) data. The FCC only publishes very coarse-grained data, only revealing total household adoption rates in 20% buckets (0-20%, 20-40%, etc.) for census tracts across all technologies, even though the FCC collects detailed subscriber counts from providers. The Census Bureau's American Community Survey (ACS) samples communities; for smaller geographic units, adoption rates are five-year averages and have only been available at the census tract level since 2017. It also does not indicate the access technology used nor the speed. As noted earlier, there is no fine-grained pricing information, so it is, for example, difficult to determine if the lack of affordable broadband options explains low adoption in a particular census tract.

There is also significant debate as to whether access via smartphones or shared computers, e.g., in a library or community center, should be counted as broadband adoption and use.

Given the limited data, correlating adoption with socio-demographic factors becomes very challenging and prone to errors.

We will discuss related questions in the Digital Inclusion section.

## Spectrum Economics

The cost and availability of spectrum help determine, for example, whether fixed wireless services, whether offered by WISPs or as a service riding on 4G and 5G cellular networks, can be economical and effective. Since most spectrum bands are currently in use for existing services, sharing and neighbor-band interference have become key research topics.

---

[22] Boris Houenou and Steve Lanning, "Survey of the US Small and Medium Providers in the Residential Broadband Internet: New Sampling Approaches," (February 1, 2021). TPRC48: The 48th Research Conference on Communication, Information and Internet Policy, Available at SSRN: https://ssrn.com/abstract=3749620 or http://dx.doi.org/10.2139/ssrn.3749620.

[23] Andre Boik, and Hidenori Takahashi. 2020. "Fighting Bundles: The Effects of Competition on Second-Degree Price Discrimination." *American Economic Journal: Microeconomics*, 12 (1): 156-87. DOI: 10.1257/mic.20180303.

[24] ibid.

The discussion of spectrum economics centered on the need for new methodologies for managing interference. These need to be risk-informed, but quantifying risk remains challenging and contentious. While interference measurement is usually in the realm of radio engineering, a better understanding of the economic impact of risks, such as the cost of degraded quality or intermittent availability, and risk mitigation, such as temporarily using a different band or higher-quality RF filters, would help inform discussions on spectrum sharing and use of adjacent spectrum bands.

The attendees stressed the social value of spectrum holdings and usage in the longer term beyond the fact that a spectrum resource is available ("is it turned on?"). They also remarked that the usual cost evaluation metrics based on $/MHz/pop (dollars per MHz of spectrum and population) is increasingly unhelpful as different spectrum bands allow different spatial reuse, for example. Characterizing the benefit of incremental licensed-by-rule ("unlicensed") spectrum remains contested and subject to wildly divergent estimates. Also, new applications such as IoT and connected vehicles may use spectrum more dynamically or only sporadically, but need coverage even in areas where nobody or few people live.

**RQ:** How can we characterize the risk of spectrum interference and the cost of mitigation or avoidance?

**RQ:** How can we measure or estimate the social and economic value of spectrum, both for traditional cellular voice and data services, as well as for services that do not pay for data or access, such as unlicensed use (Wi-Fi or industrial, scientific and medical use), radar systems or global navigation satellite systems?

**RQ:** What novel spectrum management approaches can lower cost of operations and spectrum acquisition, as well as improve performance, reliability, geographic reach, and temporal availability? What lessons can we learn from CBRS (3.5 GHz) or TV white spaces?

## Measuring and Improving Impact: Labor Market, Health, Economic Mobility

The discussion so far has focused on the technology and economics of deploying broadband, but gaining an understanding of its impact can also help motivate and focus policies — or yield more realistic expectations.

The pandemic has served as an unwelcome natural experiment illustrating the range of labor market, health and economic mobility impacts. Without doubt, the use of broadband for remote work and remote learning had the largest impact. While "working from home" and telecommuting has been discussed for over twenty years, telecommuting had previously been mostly a perk, with a self-selected set of participants. Professionals generally had broadband at home, so telecommuting was often an additional perk to avoid a trip to the office on Fridays (or a burden, to answer email after hours after a day at the office). However, some freelance and at-home call center jobs had already made broadband a job enabler. The pandemic changed telecommuting into the only form of commuting for many white collar workers, but left those without reliable internet access disadvantaged. There has been discussion that remote work allows employees to relocate to cheaper or more desirable locations, or just be closer to family. But we have a limited understanding whether such moves are anecdotes, or the beginning of a new mode of working at scale. Similarly, it is unclear whether rural locations beyond the typical commuting radius can attract such remote work or remote workers longer term.

In general, despite several attempts, quantifying the economic impact of broadband deployments has proven to be difficult. Often, the mechanism for any gains in employment or income are poorly understood. For example, when new broadband-intensive businesses move into areas that acquire or improve broadband, do they retain (younger) customers or are they able to attract more and higher-qualified workers? Or do more workers pick up additional remote work? Also, areas that have become attractive places to live or locate businesses in for other reasons are likely to attract new housing developments and thus most likely better broadband, making cause and effect hard to distinguish.

More broadly, the impact on industries and trade remain unclear. For example, does better broadband allow local firms to reach a larger market or a wider set of suppliers?[25] Does widespread internet adoption expose smaller businesses, such as retail, small-scale contract manufacturing, and tradable services, to more competition?

**RQ:** How many employees and self-employed workers rely on broadband to earn their living?

**RQ:** How are remote work and hybrid (telework) going to change where businesses are located and where their employees live?

**RQ:** What are the mechanisms by which broadband access or improvements in speed or reliability improve the economic competitiveness or standard of living, and can such changes be separated from other developments?

**RQ:** Are there measurable impacts on specific industries and institutions that either compete with or could benefit from broadband availability and adoption, such as local news media, cultural institutions, or movie theaters?

# Digital Inclusion

Full digital inclusion implies that every person who wants it has access to internet services and communication, including 24-hour daily access, quality and reliability of connectivity, security, and affordability. Access should be defined to include usability and usefulness for attaining the necessities of modern life such as food, banking, and supplies; to communicate with family, friends, co-workers, and professional service providers; to obtain health information and healthcare and local, regional, state, and federal government services (including election and voting information); to learn about job opportunities, apply for work, and do remote work; to improve one's education, training, and career mobility; to give and receive aid before, during, and after emergencies and to have business connectivity with suppliers and customers, as well as the opportunity to expand a business and find business partners.

Research and investigative reporting throughout the pandemic have demonstrated that a considerable minority of U.S. residents lack sufficient broadband or other forms of internet access, creating a modern digital divide. Disadvantaged populations without sufficient access include many rural residents, low income or unemployed people, people with high loads of debt or responsibility for others' support, people lacking digital literacy, vulnerable older people, people in substandard homes or lacking any home at all, children living in crowded homes or lacking in usable devices, disabled people, and many prisoners and others in institutionalized settings.

---

[25] Anders Akerman, Edwin Leuven, and Magne Mogstad, "Information Frictions, Internet, and the Relationship between Distance and Trade," American Economics Journal: Applied Economics (forthcoming, 2021). Available at https://www.aeaweb.org/articles?id=10.1257/app.20190589.

Predicting lack of access, understanding its causes, and tackling the digital divide(s) have been hampered by the fact that most studies are small and not generalizable, or they focus on just one dimension of access and neglect others. We lack sufficient research to understand how access is impaired in all parts of the interdependent IT system that, for digital inclusion, requires affordability, reliability, usability, and usefulness. Insufficient research exists on the causal or moderating influence of social, economic, and other factors that appear correlated with lack of access for individuals, families, businesses, and communities. Evidence-driven policy requires knowledge of the true impact of the digital divide. Some recent work suggests that the influence of poor or unreliable internet access on individual lives and communities may have been greatly underestimated, and that a lack of full internet access for some populations and groups has led to or fostered lower competition, discouraged business formation, and even led to small business failures, un- or under-employment or unemployability, or school dropouts. It likely has reduced achievement, lowered college attendance, prevented timely access to health care and caused poorer health. Lack of internet access also makes it more difficult for individuals and families to benefit from social services such as SNAP or file for unemployment insurance. A commitment to research on digital inclusion could lead not only to better fundamental knowledge of how our economy and society is functioning for all, but also whether, and how, internet access has become a necessary infrastructure of our society, much as roads, bridges, and electricity became essential in years past.

**Overall RQ:** What are the causes and impacts of digital inclusion and exclusion that will lead to a much improved fundamental understanding of the role of broadband and associated technologies and services in people's lives and life opportunities? Is there a stronger, evidence basis for policy and decision making, and more human-centric design of new broadband and related technologies?

## Adoption: Availability, Affordability, and Context

Policy makers are often interested in why households fail to "adopt" new technologies, even those that are available and offer enhanced efficiency and quality of life (such as air conditioning or online banking). Surveys can be inconclusive or misleading. For example, non-adopters cannot answer questions about access they have not experienced, so their responses such as "I don't need it" or "price too high" may only represent an attempt to provide a noncommittal answer. Ethnographic research suggests that families may consider broadband last among the expenses they can afford, e.g., after rent, food, medical, utilities (electric and gas) and car expenses. Thus, low-income households understand the value of broadband, e.g., as a way to save time,[26] but cannot afford it.[27] They may lack devices, privacy, places to store devices, or basic information about how to use browsers, make online payments, and search. They may be experiencing stress and conflict in the household, with no safe place to interact online. They may lack knowledge about how to evaluate information or distrust online information. Further, decisions to adopt broadband may vary over time with a family's changing economic conditions, a move to a new location, or the loss of a device. Hence, family decisions, such as they are, need to be tracked over context and time.

For some households, the lack of affordable broadband of sufficient quality explains the lack of adoption. Even though researching broadband availability appears relatively straightforward, with generalizable insights ("urban areas have access to high-speed broadband; rural areas often do not"), these generalizations increasingly hide more than they reveal. For example, many rural areas served by electric and telephone co-ops may have access to more affordable higher-speed fiber-to-the-home (FTTH) service than residents of dense urban areas or inner suburbs. Deployment is largely governed by decisions based on business cases of both the dominant large carriers as well as thousands of smaller providers largely serving rural areas. For areas where incumbent local exchange carriers do not see a business case for upgrading DSL to FTTN or FTTH and which are beyond the service reach of cable operators, deployment largely depends on whether smaller local carriers or new entrants can obtain government funding through the FCC Universal Service Fund (USF) or the US Department of Agriculture Rural Utility Service, e.g., through the ReConnect program. However, applying for such grants requires significant organizational capacity to meet the filing requirements. Thus, as the easier-to-serve areas and areas with well-established local providers obtain broadband, communities without such resources may be left behind or may have to rely on higher-priced satellite services.

Areas not eligible for government subsidies may suffer from digital redlining[28], where local providers decide that upgrading speed or reliability of the last-mile infrastructure is not profitable due to the perceived level of interest or ability to pay. Prieger defines redlining as situations "where broadband carriers avoid areas with high concentrations of poor and minority households"[29].

Such urban areas are generally ineligible for government subsidies as the census block is nominally served at 25/3 Mb/s – even if parts of the block are not served, network reliability is poor and the actual performance falls short of the advertised speed, possibly because landlords may not want to upgrade internal wiring or refuse to connect to additional providers.

However, beyond basic population density metrics, the availability of broadband services is poorly understood. Population density only explains a small part of the variation in broadband availability, particularly within the broad expanse of the United States categorized as rural. There are, for example, indications that winter temperatures, homes per road mile and legal prohibitions or restrictions preventing community broadband may all be partially explanatory variables. Without a better understanding of these variables, government broadband subsidies may well further exacerbate existing inequalities or, at least, spend government funds on areas that least need it. Unfortunately, the multitude of plausible explanatory variables makes studying this crucial problem difficult.

---

[26] John Horrigan, " Access and Impacts: Exploring how internet access at home and online train- ing shape people's online behavior and perspectives about their lives, " Technology Policy Institute, May 2021, available at https://techpolicyinstitute.org/wp-content/uploads/2021/05/HorriganIE.pdf.

[27] See, for example, Colin Rhinesmith, Bianca Reisdorf and Madison Bishop, "The ability to pay for broadband," *Communication Research and Practice,* 5 (2), 2019, DOI 10.1080/22041451.2019.1601491.

[28] See for example [ATT Redlining Release (pdf)](ATT Redlining Release (pdf)).

[29] Prieger, James E., The Supply Side of the Digital Divide: Is There Redlining in the Broadband Internet Access Market? (December 2001). Available at SSRN: https://ssrn.com/abstract=297499 or http://dx.doi.org/10.2139/ssrn.297499

The existing broadband deployment data captured via FCC Form 477 is known to be rather optimistic in its coverage estimates, as any block where even *one* household is served or could be served at a particular speed is counted as served at that speed. Speed claims are based on carrier filings, not performance measurements, even as DSL, satellite and fixed wireless often struggle to consistently reach the advertised speed. Often, even basic statistics are contested. For example, the FCC estimates that 1.1% of households obtain their broadband service from a wireless internet service provider (WISP), while their trade association puts it at 4.6% for 2021[30].

The Broadband DATA Act (PL 116-130) signed into law in March 2020 may greatly improve the spatial granularity of broadband deployment data once implemented by the FCC through the Universal Service Administrative Company (USAC) in late 2021 or 2022. However, while all carriers must report the number of customers they actually serve in a particular region, the FCC has refused to make this data available to researchers; instead, researchers have to rely on the Census American Community Survey (ACS) but it gathers data only by sampling. The ACS data is based on household surveys and does not differentiate between speed tiers. Thus, crucial data to evaluate adoption is being gathered, but not available to researchers. Other data, such as population demographics, road miles, topography and climate are theoretically available from government and non-governmental sources, but take inordinate effort to collect from disparate sources, and then to update, clean and analyze. In many cases, longitudinal data is needed to provide historical context. For example, coverage areas of rural electric cooperatives do not seem to be readily available. Since each effort requires inordinate amounts of preparatory work, few are undertaken. It is also nearly impossible to reproduce earlier analysis, as the precise nature and vintage of data sets is often not stated in publications.

However, given the very large number of plausible factors that can help explain broadband deployment, adoption and impact, randomized trials of policy interventions such as subsidies or training may also play a significant role in furthering our understanding of deployment and adoption, although such trials may be difficult to conduct at sufficient scale. Where policy interventions, by their nature, have an element of randomness, these interventions should be designed to be observed. (For example, not all broadband carriers may participate in the Emergency Broadband Benefit program or offer low-cost internet plans for low-income families, thus creating a natural experiment.)

---

[30] See "The 2021 Fixed-Wireless and Hybrid ISP Industry Report," WISPA and Carmel Group, 2021. Available at https://www.wispa.org/docs/2021_WISPA_Report_FINAL.pdf.

**Research Question (RQ):** What key factors, including demographic, organizational capacity, topography and legal restrictions explain or predict the availability of broadband, its performance and the likelihood that local organizations will be able to apply for grants and subsidies?

**RQ:** What kind of broadband services are available in urban, rural, and low-income communities? How do they differ from those in geographically-adjacent communities? Are they being upgraded at the same rate as those in higher-income areas?

**RQ:** What kind of natural and randomized experiments would be most useful in furthering our understanding of deployment choices, adoption and impact?

**Recommendation:** NSF should fund efforts to collect and maintain data that are critical to understanding broadband deployment and adoption, as a core broadband data repository, preferably in a cloud-based database.

**Recommendation:** Federal agencies with restricted broadband deployment or adoption data, such as the FCC, should make the data available to researchers under appropriate safeguards that protect legitimate competitive interests of providers and privacy of their customers.

**Recommendation:** NSF should work with agencies, foundations and carriers administering policies or creating programs meant to further broadband availability and adoption to include mechanisms for studying the impact and gather data that can be used for observing natural experiments.

For many people, public libraries can play a central role in providing broadband access and support. Libraries may offer a safe and comfortable place to interact with the internet and with online resources such as job applications, health information, voter registration, school homework assignments, and tax documents. Less is known about other community places that offer internet access and help--senior centers, churches, workplaces, laundromats, and so forth. A more subtle factor is community influence. When people see others like themselves using internet resources in such places, it may encourage them to do so as well. A major concern, however, is how privacy can be protected in these public spaces, and whether people learn how to safely store and recover their information over sessions.

**RQ:** Research on the causes that create and widen the digital divide in society should include contextual factors in homes and in public and community spaces. Research could improve our understanding of both the idiosyncratic and systemic contexts of adoption, such as the roles of economic mobility in the community, inadequate schooling, home ownership, and geographic segregation or isolation.

## Measuring and Improving Impact: Education

The COVID pandemic threw school districts into crisis mode as they attempted to deliver virtual education. That experience[31] showed that many students, even in areas with good broadband availability, were unable to participate in classes, whether due to lack of affordable broadband, lack of appropriate devices, insufficient technical support, or simply lack of quiet space at home or the need to earn money for the family. Most people hope that in-person instruction will be the default mode of education, but broadband has now become critical to complete homework, to communicate with parents and to find post-secondary opportunities (leading to the "homework gap" for those who lack reliable access to broadband internet).

---

[31] https://www.air.org/project/national-survey-public-education-s-response-covid-19

**RQ:** What factors, beyond availability, caused families to miss out on virtual instruction? Which models and interventions were most successful in engaging students and keeping them on their educational path?

## Training and Skills - Digital Literacy and Lifelong Learning

For many years, practitioners have recognized that broadband adoption relies on developing digital skills and that broadband use is more likely to improve societal outcomes if new users can develop information technology and literacy skills that improve their access to education and offer better job opportunities. However, identifying best practices and measuring the impact of digital skills training has been challenging, partially because of limited and intermittent funding, and partly because training is often still ad-hoc and outcomes are harder to measure than, say, through standardized tests. Should digital skills other than the most basic interactions be taught separately through digital literacy training or integrated into other job skill training programs?

The changing education models, including bootcamps, short-form courses, and micro-credentials, are also impacting how the general public can acquire broadband skills and use broadband to acquire new skills online relatively quickly. For example, thousands of COVID-19 contact tracers were trained in online bootcamps, and even truck drivers can learn some of their driving skills in online simulators. But community colleges need resources to develop appropriate curriculum and student support infrastructure.

It is still unclear if access to broadband plays a role in lifelong learning and in facilitating career transitions. Evidence of the effectiveness of both online learning and micro-credentials as well as their influence on career outcomes is either missing or incompletely documented (anecdotal). The existence of a continuum of lifelong learning opportunities completely based on self-progress including many graduate and undergraduate programs such as EdX and Harvard extension schools depends on broadband access. However, it is not clear if these expand access to non-traditional learners or how they effectively promote engagement in active online learning. There may be a need for longitudinal panel studies, as these have provided valuable feedback about the impact of Head Start and public health efforts especially in impoverished and marginalized neighborhoods.

**RQ:** What mechanisms and approaches make training job-related skills online successful? What skills can be taught partially online and where has online training faltered?

**RQ:** How can researchers track the long-term impact on graduates of online programs for certificates and other short-duration training, e.g., on career progression, employment stability and income?

## Subsidy and Other Adoption Programs

Both governments and many internet service providers offer income-restricted subsidies, such as the Lifeline program for phone and mobile internet service, the 2021 Emergency Broadband Benefit (EBB) or Comcast Internet Essentials[32]. The FCC's e-Rate program has subsidized in-school and some public library internet access since 1996.[33] Unfortunately, the impact of these programs is hard to determine. We do not know the demographics of recipients or, if collected, they are not made available to researchers. Underserved sub-populations such as incarcerated individuals or non-citizens fall outside the reach of many of these programs.

Many of these programs also suffer from poor adoption. For example, Lifeline, offering a $9.25 subsidy primarily for mobile voice and data services, has participation rates as low as 2% (Montana) and as high as 68% (Puerto Rico),[34] with little understanding of why. Is it because they are poorly advertised, with few or no carriers, and people do not know about them? Why do people stop using them? There have been proposals to integrate government support for broadband connectivity with other programs that confer eligibility, thus increasing the number of carriers willing to offer such services and reducing the administrative burden on beneficiaries and carriers. (Currently, most major wireless carriers do not participate in the Lifeline program, leaving enrollment to specialized MVNOs.) For example, the SNAP card could automatically serve as payment for subsidized broadband service since every SNAP recipient is eligible for the Lifeline benefit.

**RQ:** What are the demographics of recipients of subsidy programs?

**RQ:** What social factors predict whether a household avails themselves to broadband discounts or Lifeline subsidies? Which policies and outreach mechanisms increase participation?

**RQ:** What indicators of local capacity predict whether a school district, rural health care facility or carrier participates in the FCC Universal Service programs?

**RQ:** Do broadband subsidy programs improve long-term internet adoption, employment and health?

## Impact: Government and Nonprofit Services

Compared to even five years ago, households without broadband access have a much harder time accessing government services at the federal, state and local level. Indeed, during the pandemic, many such services, whether registering for unemployment benefits or signing up for COVID-19 vaccinations, were essentially only accessible via the internet, with overloaded call centers as the only alternative.

Thus, universal broadband access, whether through home internet access, smartphones (as long as government services are accessible via small screens), libraries or other shared

---

[32] See, for example, George W. Zuo, "Wired and Hired: Employment Effects of Subsidized Broadband Internet for Low-Income Americans," *American Economic Journal: Economic Policy* (forthcoming, 2021). Available at https://www.aeaweb.org/articles?id=10.1257/pol.20190648.
[33] https://www.fcc.gov/general/e-rate-and-education-history
[34] https://www.usac.org/lifeline/learn/program-data/

resources, will increasingly be required to provide equitable access to core government services.

**RQ:** How are households without reliable in-home internet access disadvantaged in their access to government services beyond education?

**RQ:** What are successful models to leverage shared internet access, e.g., in libraries, senior centers, homeless shelters or low-income housing?

**RQ:** What are barriers to effective use, beyond connectivity, and how do these differ between different community demographics?

## Impact: Building Community

Much of the recent Digital Inclusion research has focused on individual enablement and its related economic impact. We do not understand the impact of broadband on communities more broadly, on their cohesion and collective ability to initiate or effect change.

Broadband availability, adoption and digital resources can be tools for online activism, but also for building and maintaining communities locally and across wider geographies, as well as across demographics and age groups. Internet access allows asynchronous communication with, for example, mailing lists, Facebook groups and online collaboration tools. However, the outcomes of broadband usage are probably most ambivalent in this area, as these tools can also foster division, spread misinformation and disinformation, sow hate and discourage participation in community activities.

The participants also mentioned that there is little information on how the pandemic strengthened or reduced civic engagement, particularly at the local level. For example, many local governments started live-streaming their council and board of education meetings live, theoretically making it easier for residents to listen or participate. Many members of Congress started holding virtual town halls, again possibly increasing ways for community members to engage with their representatives.

Even services that have traditionally allowed those without internet access to stay connected to their communities and obtain local information increasingly depend on broadband access. For example, during the COVID-19 pandemic, many communities relied on public, educational and governmental (PEG) access operations of local cable systems to broadcast everything from local government meetings to school graduations and health information to their communities, but the PEG staff and volunteers sometimes struggled to get reliable internet access to produce content, stream meetings and events, and train community members in new digital tools[35].

**RQ:** What are examples of broadband and digital tools facilitating or inhibiting community organization?

**RQ:** Does broadband availability affect civic participation, such as voting and attending or contributing to local government meetings or town halls?

---

[35] See Patricia Aufderheide, Antoine Haywood and Mariana Sanchez Santos, "PEG Access Media: Local Communication Hubs in a Pandemic," Center for Media & Social Impact, American University, August 2020. Available at https://cmsimpact.org/report/peg/.

# Conclusion

This report summarized the outcomes of the November 2020 Workshops on Broadband Research, when for three consecutive weeks a total of more than 60 participants addressed the challenges of broadband and defined research questions and resources to address them across the intertwined dimensions of technology, economics and digital inclusion.

The participants agreed that broadband research deserves to be recognized as an area of high impact on policy at the federal, state and local level, though challenges range from uneven availability of data to its existence at the intersection of both engineering, economic and social sciences, and policy evaluation. Answering the research questions posed here can address one of the great challenges affecting both urban and rural residents of the United States -- making it possible for everybody to effectively use the internet to improve their education, health care and job prospects and to become a more engaged citizen.

# Living Bibliography

Relevant publications on the different the workshop topics are available here:

https://docs.google.com/document/d/1fgjv-RrYH3NTKnzP8My4o_4IuK0lBVIrb48Q9NqZHeQ/edit#

# Appendix

## Workshop Participants

Affiliations are listed for identification purposes only and do not imply endorsement of positions or this report by these organizations. All government employees attended in their personal capacity.

- Ian Akyildiz (Georgia Institute of Technology)
- Moussa Arab (Dubai)
- Patricia Aufderheide (American University)
- Jenith L. Banluta (School of Engineering and Architecture Ateneo de Davao University)
- Brigid Barron (Stanford University)
- Johannes M. Bauer (Michigan State University)
- Scott Bradner (Harvard University)
- Jack Brassil (Princeton University)
- Doug Brake (Information Technology and Innovation Foundation)
- Jenna Burrell (University of California Berkeley)
- Octavian Carare (FCC)
- Trina Coleman (Virginia Union University)
- Shelia R Cotten (Clemson University)
- Udayan Das (Loyola University Chicago)

- Nick Feamster (University of Chicago)
- Harold Feld (Public Knowledge)
- Nick Feamster (University of Chicago)
- Kenneth Flamm (University of Texas at Austin)
- Rob Frieden (Penn State University)
- Peter Joseph Gloviczki (Coker University)
- Rafi Goldberg (NTIA)
- Amy Gonzales (University of California Santa Barbara)
- Karen Hanson (NTIA)
- Cynthia Hood (Illinois Institute of Technology)
- John Horrigan (Technology Policy Institute)
- Steven Howland (Federal Reserve)
- Peng Hu (University of Waterloo)
- Heather E. Hudson (University of Alaska Anchorage)
- Sherif Mohamed Ismail (Makkah Almokramh)
- Sonia Jorge (World Wide Web Foundation)
- Padma Krishnaswamy (FCC)
- Dan Kilper (University of Arizona)
- T.V. Lakshman (Nokia Bell Laboratories)
- William Lehr (Massachusetts Institute of Technology)
- Linda Loubert (Morgan State University)
- José Ricardo López-Robles (Autonomous University of Zacatecas)
- Tarun Mangla (University of Chicago)
- Jacob Manlove (Arkansas State University)
- Scott Marcus (Bruegel)
- Mike Martin (Census Bureau)
- Nicole P. Marwell (University of Chicago)
- Nicolas Merveille (University of Quebec in Montreal)
- Traci Morris (Arizona State University)
- Karen Mossberger (Arizona State University)
- Shannon M. Oltmann (University of Kentucky)
- Edward Oughton (George Mason University)
- Abhishek Pandey (Scsvmv University, Kanchipuram)
- Jon M. Peha (Carnegie Mellon University)
- Gopika Premsankar (Aalto University)
- K. K. Ramakrishnan (University of California Riverside)
- Anna Read (The Pew Charitable Trusts)
- David Reed (University of Colorado at Boulder)
- Bibi Reisdorf (UNC Charlotte)
- Jennifer Rexford (Princeton University)
- Colin Rhinesmith (Simmons University)
- Steve Rosenberg (FCC)
- Carolina Rossini (Portulans Institute)
- Steve Sawyer (Syracuse University)

- Alexis Schrubbe (University of Texas)
- Subhabrata Sen (AT&T Research)
- Karen Sollins (MIT)
- William Staples (University of Kansas)
- Sharon Strover (University of Texas at Austin)
- Richard Taylor (Pennsylvania State University)
- Emy Tseng (NTIA)
- Morgan Vigil-Hayes (Northern Arizona University)
- Ann Von Lehmen (National Science Foundation)
- Antwuan Wallace (National Innovation Service)
- Scott Wallsten (Technology Policy Institute)
- Stacey Wedlake (University of Washington)
- Martin B.H. Weiss (University of Pittsburgh)
- Glenn Woroch (University of California Berkeley)
- Mo Xiao (University of Arizona)
- Christopher Yoo (University of Pennsylvania)
- Matthew J Zagaja (Metropolitan Area Planning Council)
- Hongwei Zhang (Iowa State University)